\documentclass[final]{dmtcs-episciences}
\usepackage{amsmath,amssymb,color,epsfig,bbm,tikz, enumerate,lineno}
\usepackage[round]{natbib}
\usepackage{eurosym}
\newtheorem{thm}{Theorem}
\newtheorem{prop}[thm]{Proposition}
\newtheorem{lem}[thm]{Lemma}
\newtheorem{cor}[thm]{Corollary}

\newtheorem{obs}[thm]{Observation}

\author{Christopher Duffy\affiliationmark{1} \and Gary MacGillivray\affiliationmark{2}\thanks{This author acknowledges support from NSERC.} \and \'{E}ric Sopena\affiliationmark{3}}

\title{A study of~\begin{math}k\end{math}-dipath colourings of oriented graphs}

\affiliation{
	Department of Mathematics and Statistics, University of Saskatchewan, Saskatoon, CANADA\\
	Department of Mathematics and Statistics, University of Victoria, Victoria, CANADA\\
	Univ. Bordeaux,  Bordeaux INP, CNRS, LaBRI, UMR5800, F-33400, Talence, FRANCE}

\received{2016-6-27}
\revised{2017-8-3}
\accepted{2018-1-13}

\begin{document}

\publicationdetails{20}{2018}{1}{6}{1520}
\maketitle

\begin{abstract}
We examine $t$-colourings of oriented graphs in which, for a fixed integer~\begin{math}
k \geq 1
\end{math}, vertices 
joined by a directed path of length at most $k$ must be assigned different colours. 
A homomorphism model that extends the ideas of Sherk for the case $k=2$ is described. 
Dichotomy theorems for the complexity of the problem of 
deciding, for fixed $k$ and $t$, whether  there exists such a $t$-colouring are proved.
\end{abstract}

\section{Introduction}
Recall that an \emph{oriented graph} is a digraph obtained from a simple, undirected graph by giving each edge one of its two possible orientations.  Recall, also, that if $G$ and $H$ are oriented graphs, then a \emph{homomorphism of $G$ to $H$} is a function $\phi$
from the vertices of $G$ to the vertices of $H$ such that~\begin{math}
\phi(x)\phi(y) \in E(H) \end{math} whenever~\begin{math} xy \in E(G)\end{math}. If $G$ and $H$ are oriented graphs such that there is a homomorphism $\phi$ of $G$ to $H$, then we write~\begin{math}
\phi: G \to H
\end{math}, or~\begin{math}
G \to H
\end{math} if the name of the function $\phi$ is not important.

\begin{figure} 
	\begin{center}
		\includegraphics[width=0.5\textwidth]{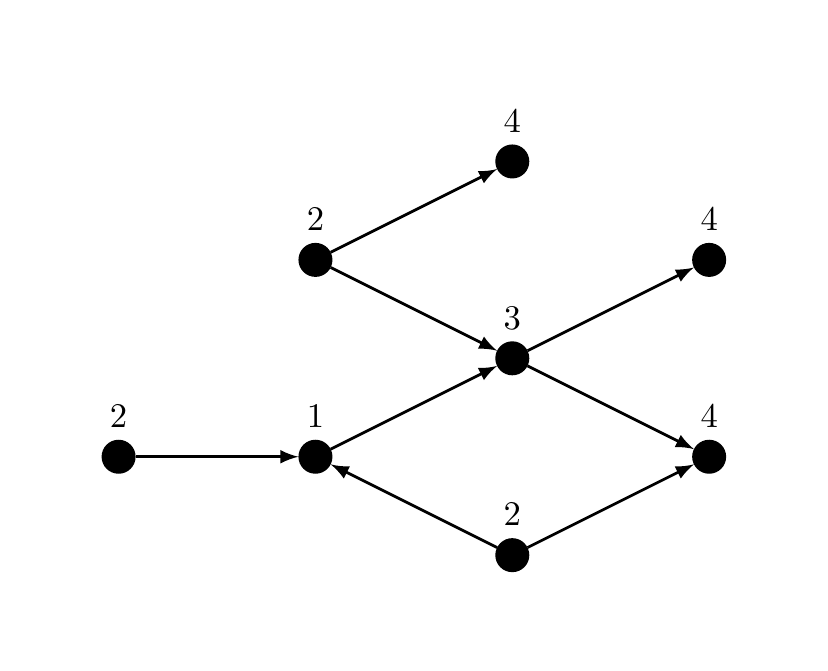}
	\end{center}
	\caption{A $3$-dipath $4$-colouring.}
	\label{kdipath:example1}
\end{figure}

Let $k$ and $t$ be positive integers, and let $G$ be an oriented graph.  \cite{MIWE06} defined a \emph{$k$-dipath $t$-colouring} of $G$  to be an assignment of $t$ colours to the vertices of $G$ so that any two vertices joined by a directed path of length at most $k$ are assigned different colours. 
A $1$-dipath $t$-colouring of an oriented graph $G$ is a $t$-colouring of the underlying undirected graph of $G$.
See Figure \ref{kdipath:example1} for an example of a $3$-dipath $4$-colouring of an oriented graph. The \emph{$k$-dipath chromatic number of $G$}, denoted by~\begin{math}
\chi_{k\text{-dip}}(G)
\end{math}, is the smallest positive integer $t$ such that there exists a $k$-dipath $t$-colouring of $G$. \cite{MIWE06}  showed that any orientation of a Halin graph has $2$-dipath chromatic number at most $7$, and there are infinitely many such graphs $G$ with~\begin{math}
\chi_{2\text{-dip}}(G) = 7
\end{math}.

For a positive integer $t$, an \emph{oriented $t$-colouring} of an oriented graph $G$ is a homomorphism of $G$ to some oriented graph on $t$ vertices. Oriented colourings were first introduced by \cite{CO94}, and have been a topic of considerable interest in the literature since then;  see the recent survey by  \cite{SO15}, and also work by \cite{BO99}, \cite{HDS06b}, \cite{SW13} for related topics. 
The $2$-dipath chromatic number is of interest, in part, because it gives a lower bound for the \emph{oriented chromatic number}~\begin{math}
\chi_o(G)
\end{math} -- the smallest positive integer $t$ such that $G$ admits an oriented $t$-colouring.
Since oriented graphs have no directed cycles of length two, the definition implies that any two vertices of $G$ joined by a directed path of length at most two are assigned different colours (\textit{i.e.}, they have different images) in an oriented colouring of $G$.  
It follows from the definition that any oriented colouring of $G$ is a $2$-dipath colouring of $G$; hence~\begin{math}
\chi_{2\text{-dip}}(G) \leq \chi_o(G).
\end{math}

In her Master's thesis \cite{KMY09} gives a \emph{homomorphism model} for $2$-dipath $t$-colouring. For each positive integer $t$, she describes an oriented graph $\mathcal{G}_t$ with the property that an oriented graph $G$ has a $2$-dipath $t$-colouring if and only if there is a homomorphism of $G$ to $\mathcal{G}_t$.
As is common with such theorems, it is possible to use the homomorphism to $\mathcal{G}_t$
to find a $2$-dipath $t$-colouring of $G$: the colour assigned to a vertex is determined by
its image (but is not equal to it).
The existence of this model implies an upper bound for the oriented chromatic number as a function of the $2$-dipath chromatic number.  It also leads to a proof that deciding whether a given oriented graph has a $2$-dipath $t$-colouring is Polynomial if the fixed integer~\begin{math}
t \leq 2
\end{math}, and NP-complete if~\begin{math}
t \geq 3
\end{math}.

A natural question is whether Sherk's results can be generalized to $k$-dipath $t$-colouring. 
We seek a model similar to hers, where the homomorphism to the target oriented graph can be used to 
find a $k$-dipath $t$-colouring of the given oriented graph.
In particular, vertices of the given oriented graph $G$ with the same image should 
be assigned the same colour.
We suggest that such model will likely exist only for colouring oriented graphs with no directed cycles of 
length $k$ or less. 
Consider the case of $3$-dipath $t$-colouring, where~\begin{math} t \geq 3 \end{math}. Suppose there exists a digraph $H_{3, t}$ with the property that an oriented graph $G$ has a $3$-dipath $t$-colouring if and only if there is a homomorphism of $G$ to $H_{3, t}$. The digraph $H_{3, t}$ has no loops, otherwise all vertices of $G$ can be assigned the same colour. 

By definition, the directed $3$-cycle has a $3$-dipath $3$-colouring: assign each vertex a different colour. Thus, there is a homomorphism of the directed $3$-cycle to $H_{3, t}$. Consequently, $H_{3, t}$ has a directed $3$-cycle.
But, now there  is a homomorphism of a directed path of length three to $H_{3, t}$ in which the two end vertices have the same image.  Since the ends of a directed path of length 3 must be assigned different colours in 
a $3$-dipath $t$-colouring, this model will not have the desired property.
Similar considerations apply to $k$-dipath $t$-colouring for all pairs of positive integers $k$ and $t$ with~\begin{math} t \geq k \end{math}. Hence, a homomorphism model of the type we seek will not exist if the oriented graphs being coloured can have directed cycles of length $k$ or less. Finally, we note that Sherk's homomorphism model for $2$-dipath $t$-colouring is for oriented graphs. These have no directed cycles of length two or less.

The main result of this paper is the construction of a homomorphism model for $k$-dipath $t$-colouring of oriented graphs with no directed cycles of length $k$ or less. That is, for all positive integers $k$ and $t$ we describe an oriented graph~\begin{math}\mathcal{G}_{k,  t}\end{math} with the property that an oriented graph $G$, with no directed cycle of length at most $k$, has a $k$-dipath  $t$-colouring if and only if $G$ admits a homomorphism to~\begin{math}\mathcal{G}_{k,  t}\end{math} and, further, 
the homomorphism can be used to find a $k$-dipath $t$-colouring of $G$. 

After presenting this result in Section \ref{kdipath:homsec},  in Section \ref{kdipath:comsec} we determine the complexity of deciding the existence of a $k$-dipath $t$-colouring  for all pairs of fixed positive integers $k$ and $t$.  When instances are restricted to oriented graphs with no directed cycles of length $k$ or less, it is shown that that this problem is NP-complete whenever ~\begin{math}
t>k\geq 3
\end{math}, and Polynomial if~\begin{math} t = k \end{math} or~\begin{math} k \leq 2 \end{math}.  When there are no restrictions, it is shown that this problem is NP-complete whenever~\begin{math} k \geq 3 \end{math} and~\begin{math} t \geq 3 \end{math}, and Polynomial whenever~\begin{math} k \leq 2 \end{math} or~\begin{math} t \leq 2 \end{math}.

\section{Preliminaries}\label{kdipath:presec}

In this section we review relevant definitions, the homomorphism model for $2$-dipath colourings, and make small improvements  to the known upper bounds on the oriented chromatic number of oriented graphs with $2$-dipath chromatic number $3$ or $4$. We also observe some straightforward extensions to $k$-dipath colouring of known results for $2$-dipath colouring. 

Let $G$ be an oriented graph, and let~\begin{math} x, y \in V(G)\end{math}.
A \emph{directed walk} is a sequence of vertices ~\begin{math}W = v_0, v_1,v_2, \dots, v_{\ell-1}, v_\ell \end{math}, such that
\begin{math}v_iv_{i+1} \in E(G)\end{math} for~\begin{math} = 0, 1, \ldots, \ell -1\end{math}.
The integer $\ell$ is the \emph{length} of $W$.
If~\begin{math}v_0 = x \end{math} and~\begin{math}v_\ell = y\end{math}, then $W$ is a \emph{directed walk from $x$ to $y$}.
Note that the vertices belonging to $W$ need not be different.
If no two vertices of $W$ are the same, then $W$ is a \emph{directed path}.
If all vertices of $W$ are different except $v_0$ and $v_\ell$, then $W$ is a \emph{directed cycle}.

If there is a directed walk from $x$ to $y$, then the vertex $y$ is said to be \emph{reachable} from $x$.
The \emph{distance} from $x$ to $y$ is defined to be the smallest length of a directed walk from $x$ to $y$, or infinity of no 
such walk exists.
The \emph{weak distance between $x$ and $y$}, denoted $d_{weak}(x,y)$, is 
defined to be the minimum of the distance from $x$ to $y$ and the distance from $y$ to $x$.

This parameter is $\infty$ if neither of $x$ and $y$ is reachable from the other.

The \emph{weak diameter} of an oriented graph $G$ is the maximum of the weak distance between any
two distinct vertices of $G$. 
A directed graph is called \emph{weakly connected} if its weak diameter is finite.

Let $G$ be a directed graph, and~\begin{math}x \in V(G)\end{math}.
The \emph{out-neighbourhood} of $x$ is~\begin{math}{N^+(x) = \{y: xy \in E(G)\}}\end{math},
and the \emph{in-neighbourhood} of $x$ is~\begin{math}N^-(x) = \{y: yx \in E(G)\}\end{math}.
The vertex $x$  is a \emph{source} if~\begin{math}N^-(x) = \emptyset\end{math}, 
and is a \emph{sink} if~\begin{math}N^+(x) = \emptyset \end{math}.
A \emph{universal source} is a source such that~\begin{math}N^+(x) = V(G) - \{x\}\end{math},
and a \emph{universal sink} is a sink such that~\begin{math}N^-(x) = V(G) - \{x\}\end{math}.
More generally, the \emph{$\ell$-out-neighbourhood} of $x$ is the set of all vertices reachable from $x$ by a directed
walk of length at most $\ell$, and the 
 \emph{$\ell$-in-neighbourhood} of $x$ is the set of all vertices which can reach $x$ by a directed
walk of length at most $\ell$.

The \emph{directed girth} of a directed graph $H$ is defined to be the minimum length of a directed cycle in $H$,
or infinity if $H$ has no directed cycle.  
Our homomorphism model for $k$-dipath colouring applies only to the family of oriented graphs
with directed girth at least $k+1$.

Let $\mathcal{F}$ be a family of oriented graphs.
The \emph{oriented chromatic number of $\mathcal{F}$}, denoted by $\chi_o(\mathcal{F})$,
is the least integer $t$ so that~\begin{math}\chi_o(F) \leq t \end{math} for all~\begin{math}F \in \mathcal{F}\end{math}.  
We  say that an oriented graph $H$ is a \emph{universal target for $\mathcal{F}$} if~\begin{math}F \rightarrow H\end{math} for all~\begin{math}F \in \mathcal{F}\end{math}.
If $H$ is a universal target for $\mathcal{F}$, then~\begin{math}\chi_o(\mathcal{F}) \leq |V(H)|\end{math}.
For example, using the quadratic residue tournament on seven vertices as a universal target
for the family $\mathcal{O}$ of orientations of outerplanar graphs, \cite{SO97} shows~\begin{math}\chi_o(\mathcal{O}) \leq 7\end{math}.

Let $G$ be an oriented graph with directed girth at least $k+1$. 
We say $G$ is a \emph{$k$-dipath clique} if~\begin{math}\chi_{k\text{-dip}}(G) = |V(G)|\end{math}.
The terminology arises by analogy with undirected graphs, where a clique is a graph 
for which the chromatic number equals the number of vertices.

Let $G$ be an oriented graph. Define $G^k$ to be the simple graph formed from $G$ as follows:
\begin{itemize}
	\item~\begin{math}V(G^k) = V(G)\end{math}, and
	\item~\begin{math}E(G^k) = \{uv | 0 < d_{weak}(u,v) \leq k \}\end{math}. 
\end{itemize}

It follows from the definitions
that there is an equivalence between  $2$-dipath colourings of $G$ and proper vertex colourings of the simple graph $G^2$ 
(see \cite{MAYO12}).  
This equivalence extends to $k$-dipath colourings of $G$ and proper vertex colourings of $G^k$.

\begin{obs}
	If $G$ is an oriented graph with directed girth at least $k+1$, then there is a one-to-one correspondence between $k$-dipath colourings of $G$ and proper colourings of $G^k$.
\end{obs}

Using this observation we generalize a result of \cite{BDS17} for $2$-dipath cliques.
This result will be used in Section  \ref{kdipath:comsec}.

\begin{prop} [\cite{BDS17}]
	An oriented graph is a $2$-dipath clique if and only if it has weak diameter at most $2$.
\end{prop}

\begin{prop}
	Let $k\geq 2$ be an integer. An oriented graph is a $k$-dipath clique if and only if it has weak diameter at most $k$.
\end{prop}

\begin{proof}
Let $G$ be an oriented graph with directed girth at least $k+1$. 
We observe that $G^k$ is a complete graph if and only if for each pair of non-adjacent vertices, 
say $u$ and $v$, there is a directed path of length at most $k$, in some direction, between $u$ and $v$. 
Equivalently, $G$ has weak diameter at most $k$.
\end{proof}

We now review the homomorphism model for $2$-dipath colouring given by \cite{MAYO12}.
Let $t$ be a positive integer.
The oriented graph $\mathcal{G}_t$ is defined as follows.
Its vertices are $(t+1)$~-~tuples in which the position among~\begin{math}1, 2, \ldots, t \end{math} indicated by the 0-th entry is filled with the place-holder  
``$\cdot$'', and the remaining positions among $1, 2, \ldots, t$ are filled with a 0 or a 1.
\begin{displaymath}V(\mathcal{G}_t) = \{    (u_0;u_1,u_2,\dots,u_t) : u_0 \in \{1,2,\dots,t\}, u_{u_0} = \cdot, u_i \in \{0,1\} \text{ for }  1 \leq i \leq t \text{ and } i\neq u_0 \} \end{displaymath}
\begin{displaymath} E(\mathcal{G}_t) = \{    (u_0;u_1,u_2,\dots,u_t)(x_0;x_1,x_2,\dots,x_t) : u_{x_0} = 1, x_{u_0} = 0\}. \end{displaymath}
It is then proved that an oriented graph $G$ admits a homomorphism to 
$\mathcal{G}_t$ if and only if $G$ has $2$-dipath chromatic number at most $t$.
That is, $\mathcal{G}_t$ is a universal target for the family of oriented graphs 
with $2$-dipath chromatic number at most $t$.
Further, suppose~\begin{math}\phi: G \to \mathcal{G}_t \end{math}.  
For each vertex $x$, assigning colour $u_0$ to $x$ if and only if the mapping $\phi$,~\begin{math}\phi(x) = (u_0;u_1,u_2,\dots,u_t)\end{math},
is a 2-dipath $t$-colouring of $G$.
Since homomorphisms compose, 
the oriented chromatic number of $\mathcal{G}_t$ is an upper bound for the oriented chromatic number of $G$.
\cite{MRS} show that~\begin{math}\chi_o(\mathcal{G}_t) \leq 2^{t}-1\end{math}, which gives the following.

\begin{thm}  [\cite{MRS}] \label{kdipath:2bounds}
	If $G$ is an oriented graph, then 
	\begin{displaymath}\chi_{2\text{-dip}}(G) \leq \chi_o(G) \leq  2^{\chi_{2\text{-dip}}(G)} -1. \end{displaymath}
\end{thm}

The topic of universal targets for $k$-dipath colourings is considered in Section \ref{kdipath:homsec}.  
Here we  offer improvements for the cases~\begin{math}t=3,4\end{math} of Theorem \ref{kdipath:2bounds}. 

\begin{prop}
	Let~\begin{math}t\geq 2 \end{math} be an integer and let $G$ be an oriented graph with~\begin{math}\chi_{2\text{-dip}} \leq t \end{math}.
	\begin{itemize}
		\item If $t \leq 3$, then $\chi_o(G) \leq 5$.
		\item If $t \leq 4$, then $\chi_o(G) \leq 12$.				
	\end{itemize}
\end{prop}

\begin{figure}
	\begin{center}
		\includegraphics[width=0.75\textwidth]{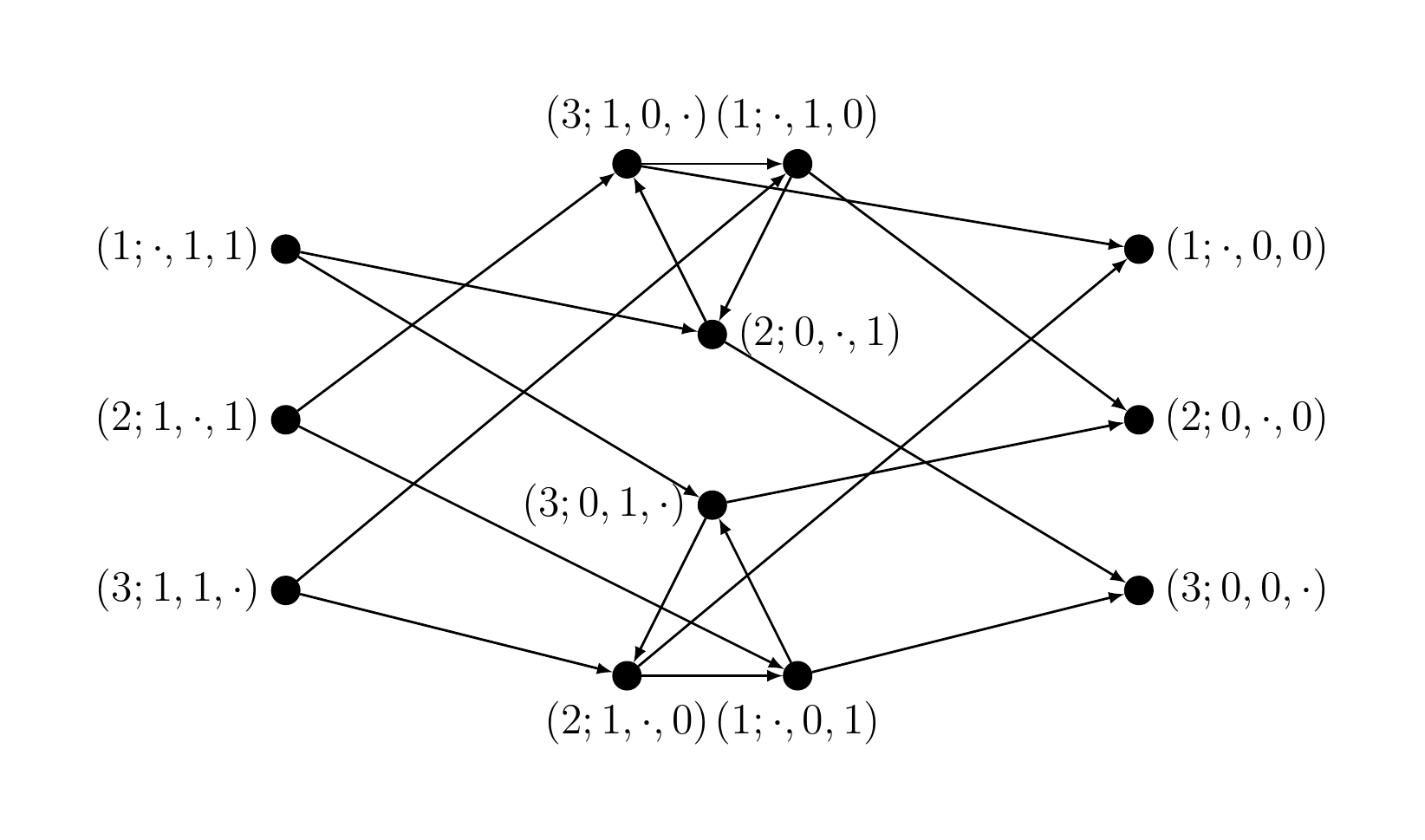}	
	\end{center}
	\caption{The universal target for the family of oriented graphs with$\chi_{2\text{-dip}} \leq 3$.}
	\label{kdipath:3dptarget}
\end{figure}

\begin{figure}
	\begin{center}
		\includegraphics[width=0.75\textwidth]{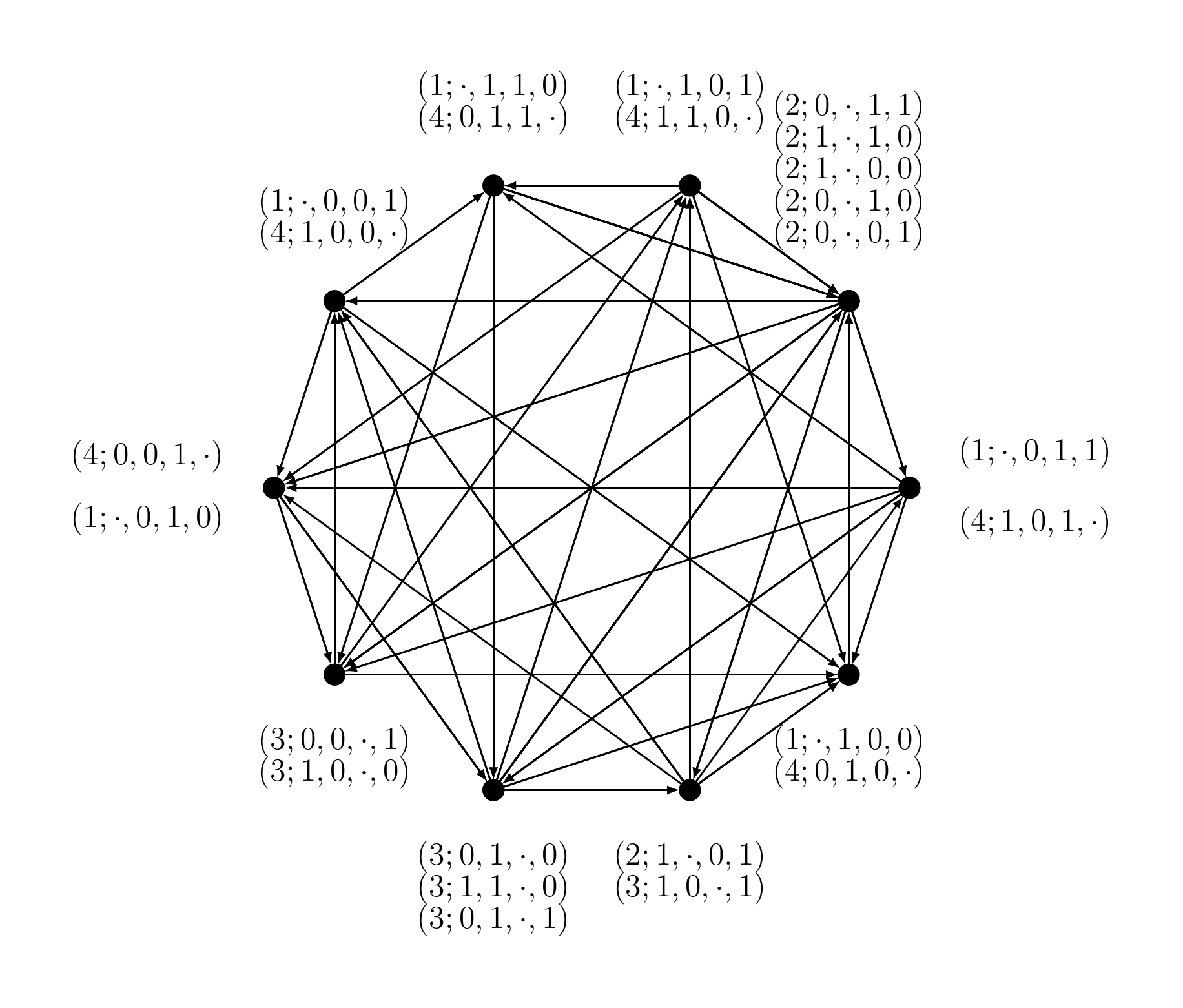}	
	\end{center}
	\caption{A homomorphic image of the modified universal target for the family of oriented graphs with $\chi_{2\text{-dip}} \leq 4$.}
	\label{kdipath:4dptarget}
\end{figure}

\begin{proof}

Figure \ref{kdipath:3dptarget} shows $\mathcal{G}_{3}$, 
except for arcs between the source vertices on the left and the sink vertices on the right.

By inspection, $\mathcal{G}_3$ admits a homomorphism to the  tournament of order $5$ formed from a 
copy of a directed $3$-cycle together with a universal source vertex  and universal sink vertex. 
This proves the first statement.

Let $H$ be he oriented graph obtained from 
$\mathcal{G}_{4}$
by deleting all sources and all sinks.
Figure  \ref{kdipath:4dptarget} gives a mapping of $H$ to the oriented graph of order $10$ shown.
Thus, this oriented graph, together with a universal source and universal sink vertex, is a homomorphic image of $\mathcal{G}_{4}$. 
This proves the second statement.
\end{proof}

For~\begin{math}k \geq 2\end{math}, a $k$-dipath colouring of an oriented graph $G$
is a $2$-dipath colouring of $G$.
Thus, Theorem \ref{kdipath:2bounds} implies the following result for the $k$-dipath chromatic number.

\begin{cor} \label{kdipath:kdpupper} 
	If $G$ is an oriented graph with directed girth at least $k+1$, then 
	\begin{displaymath}\chi_o(G) \leq 2^{\chi_{k\text{-dip}}(G)} -1.\end{displaymath}
\end{cor}

Note, however, that by Theorem \ref{TargetHoms}, $\chi_{k\text{-dip}}$ can always be replaced by $\chi_{2\text{-dip}}$. 
By contrast, the lower bound for $\chi_o(G)$ given in Theorem \ref{kdipath:2bounds} does not hold 
if $\chi_{2\text{-dip}}$ is replaced by $\chi_{k\text{-dip}}$.
The oriented chromatic number of a directed path on $4$ vertices is $3$,
but its 4-dipath chromatic number is $4$.

\section{Homomorphisms and $k$-dipath Colouring} \label{kdipath:homsec}

We begin our study of homomorphisms and $k$-dipath colourings by considering oriented graphs without directed cycles. 
The transitive tournament on $t$ vertices is denoted by $T_t$.

\begin{thm}[\cite{NP78}] \label{kdipath:acyclicmsw} 
If $G$ is an acyclic oriented  graph, then~\begin{math}G \rightarrow T_t \end{math} if and only if 
there is no homomorphism of a directed path on $t+1$ vertices to $G$.
\end{thm}

A different way to phrase the condition in the theorem is that if~\begin{math}G \rightarrow T_t\end{math}, then $G$ has no directed cycles, and no
directed path of length $t$ or more.

\begin{cor} \label{kdipath:acyclicme}
	If $G$ is an acyclic oriented graph and the longest directed path in $G$ has $t$ vertices, then~\begin{math}\chi_{k\text{-dip}}(G) = t \end{math} for all ~\begin{math}k \geq t > 1\end{math}.
\end{cor}

\begin{proof}
Suppose $k \geq t$.

Observe that since $G$ has a path on $t$ vertices, we have~\begin{math}\chi_{k\text{-dip}}(G) \geq t\end{math}. By Theorem \ref{kdipath:acyclicmsw}, there exists a homomorphism~\begin{math}\phi: G \to T_t\end{math}.
Since $T_t$ has no directed cycle, 
any two vertices joined by a directed path must have different images.
If the vertices of $T_t$ are regarded as the colours $1, 2, \ldots, t$, then 
$\phi$ is a $k$-dipath colouring of $G$.
Thus~\begin{math}\chi_{k\text{-dip}}(G) \leq t \end{math}.  
\end{proof}

Since any oriented graph which admits a homomorphism to $T_t$ is acyclic, it follows that $T_t$ is a universal target
for $k$-dipath $t$-colouring of acyclic oriented graphs.

\begin{cor} \label{kdipath:acycliccor}
Let $k$ and $t$ be positive integers. An acyclic oriented graph $G$ has~\begin{math}\chi_{k\text{-dip}}(G) \leq t\end{math} if and only if $G$ admits a homomorphism to  $T_t$.
\end{cor}

\begin{cor} 
Let $k$ and $t$ be positive integers. An acyclic oriented graph $G$ has~\begin{math} \chi_{k\text{-dip}}(G) \leq t \end{math} if and only if $G$ has no directed path on at least $t+1$ vertices.
\end{cor}

Though not a direct analogue, Corollary \ref{kdipath:acyclicme} has a similar flavour to the early results on graph colourings of   \cite{V62}, \cite{H65} , \cite{G67}, and \cite{R67}.

Observe that for $k \geq t$, the $k$-dipath chromatic number of $T_t$ equals $t$.
Thus,  Corollary \ref{kdipath:acyclicme} states that the $k$-dipath chromatic number of $T_t$ is 
an upper bound on the $k$-dipath chromatic number
of any oriented graph which admits a homomorphism to $T_t$.
The same statement holds if $T_t$ is replaced by any oriented graph with large enough directed girth.

\begin{thm} \label{kdipath:homgirth}
Let $G$  and $H$ be oriented graphs such that $H$ has directed girth at least $k+1$. 
If~\begin{math}G \rightarrow H\end{math}, then~\begin{math}\chi_{k\text{-dip}}(G) \leq \chi_{k\text{-dip}}(H)\end{math}.
\end{thm}

\begin{proof}
Suppose~\begin{math}\phi: G \rightarrow H \end{math}. 
Let $c$ be a $k$-dipath $t$-colouring of $H$.  
Let~\begin{math}{c^\prime: V(G) \rightarrow \{1,2,3,\dots,t\}}\end{math}
be defined by~\begin{math}c^\prime(v) = c(\phi(v))\end{math}, for all ~\begin{math}v \in V(G)\end{math}.  
We claim that this is a $k$-dipath $t$-colouring of $G$.

Let~\begin{math}u, v \in V(G)\end{math}. 
Suppose that there is a directed path from $u$ to $v$ of length $\ell \leq k$.
Then there is a directed walk $W$ of length $k$ in $H$ from $\phi(u)$ to $\phi(v)$.
Since the directed girth of $H$ is $k+1$ or more, $W$ is a directed path.
Thus,~\begin{math}c(\phi(u)) \neq c(\phi(v))\end{math}.
Therefore~\begin{math}c^\prime(u) \neq c^\prime(v)\end{math}, and  $c^\prime$ is a $k$-dipath colouring of $G$.
\end{proof}

We now  describe a homomorphism model for $k$-dipath $t$-colouring of oriented graphs with directed girth at least $k+1$.
That is, for all positive integers $k$ and $t$, we will describe an oriented graph $\mathcal{G}_{k,  t}$ 
with the property that an oriented graph $G$, with no directed cycle of length at most $k$,
has a $k$-dipath  $t$-colouring if and only if 
$G$ admits a homomorphism to $\mathcal{G}_{k,t}$ and, further, 
the homomorphism can be used to find
a $k$-dipath $t$-colouring of $G$.
The graph $\mathcal{G}_{k,t}$ is a universal target for $k$-dipath $t$-colouring.
Once this model is in place, Theorem \ref{kdipath:homgirth} could be viewed as a direct consequence of
the fact that a composition of homomorphisms is a homomorphism;  it arises in the proof for the model, however.

We begin by defining a special set of matrices.  Let~\begin{math}k \geq 2 \end{math} be an integer.
Suppose that an oriented graph $G$, with directed girth at least $k+1$,
admits a $k$-dipath $t$-colouring, $c$.
For~\begin{math} x \in V(G)\end{math}, given $c(x)$, we want to encode information about the colours of vertices 
in the $k$-in-neighbourhood of $x$ and $k$-out-neighbourhood of $x$.

We define the \emph{$k$-dipath colouring matrix of $x$ with respect to $c$} to be the~\begin{math}(2k-1) \times t \end{math}
zero-one matrix  
$A_{x, c}(G)$
with rows indexed by  
\begin{math}-(k-1), -(k-2), \dots,-1, 0, 1, \dots,(k-2), (k-1)\end{math}, columns indexed by~\begin{math}1, 2, 3, \dots, t \end{math}, and  $(i, j)$-entry equal to 1 
if and only if there exists a vertex~\begin{math}y \in V(G) \end{math} such that~\begin{math}c(y) = j \end{math}, and 
\begin{itemize}
	\item if~\begin{math}i \in \{-(k-1), -(k-2), \dots, -1\}\end{math}, then there is a directed path from $y$ to $x$ of  length $i$;
	\item if~\begin{math}i \in \{1, 2, \dots,  (k-1)\}\end{math}, then there is a directed path from $x$ to $y$ of length $i$; and
	\item if $i = 0$, then $x = y$.
\end{itemize}
When the graph $G$ is clear from the context, or unimportant in the discussion,
the $k$-dipath colouring matrix of $x$ with respect to $c$ is denoted by
$A_{x, c}$.

We illustrate the definition with an example. 
Consider the colouring, $c$, given in Figure \ref{kdipath:example1}. 
Let $x$ be the unique vertex such that~\begin{math}c(x) = 3 \end{math}. The $3$-dipath colouring matrix of $x$ is given by

\begin{center}
	\begin{tabular}{c|cccc}
		& $1$ & $2$ & $3$  & $4$ \\ 
		\hline $-2$ &$0$  &$1$  &$0$  & $0$ \\ 
		$-1$ & $1$  &$1$  &$0$  &$0$  \\ 
		$0$& $0$ & $0$ &$1$  &$0$  \\ 
		$1$ & $0$ & $0$  & $0$ & $1$ \\ 
		$2$ & $0$ & $0$ & $0$ & $0$ \\ 	 
	\end{tabular} 
\end{center}

For example, the value, $1$, in 
the entry~\begin{math}(-1, 2)\end{math}
arises because there is a vertex $w$ such that~\begin{math}c(w) = 2\end{math}
and there is a directed path of length $1$ from $w$ to $x$.

Let $\mathcal{A}_{k, t}$ denote the set of all possible $k$-dipath colouring matrices 
over all $k$-dipath $t$-coloured oriented graphs with directed girth at least~\begin{math}k+1\end{math}. We note that $\mathcal{A}_{k, t}$ is necessarily finite, as each member is a~\begin{math}(2k-1)\times t\end{math} matrix with entries from $\{0,1\}$.

Since vertices at weak distance at most $k$ must be assigned different colours, 
each element of $\mathcal{A}_{k, t}$ is a matrix that satisfies several conditions.
Suppose $A$ is the $k$-dipath colouring matrix of $x$ with respect to $c$. 
If the colour of $x$ is $j$, then no vertex at weak distance at most $k$ from $x$ can
be assigned colour $j$.
Thus, the entry of $A$ in column $j$ and row 0 equals 1, and all other entries in column $j$ are zero.
If there is a directed path of length $i < k$ from $y$ to $x$, then no vertex 
$w$ for which there is a directed path of length $k-i$ from $x$ to $w$ can have the same colour as $y$.
If the colour of $y$ is $j$, then the entry of $A$ in column $j$ and row $-i$ equals 1, and
 the entries of $A$ in column $j$ and rows $0, 1, \ldots, k-i$
are all equal to zero.
Note that no assertion can be made about the entries of $A$ in column $j$ and rows~\begin{math}-(i-1), -(i-2), \ldots, -1\end{math}
because there is no guarantee that a vertex $v$ for which there is a directed path of length at most $i-1$ 
to $x$ lies on the directed path from $y$ to $x$.
Similar considerations apply to vertices $w$ for which there is a directed path of length $i < k$ from $x$ to $w$.
These conditions are formalized succinctly as follows.

\begin{obs} \label{kdipath:colstruc}
Let~\begin{math}A = [a_{ij}] \in \mathcal{A}_{k, t}\end{math}.  Suppose ~\begin{math}a_{ij}=1\end{math}.
Then, 
	\begin{itemize}
		\item  if~\begin{math}i = 0 \end{math}, then~\begin{math}a_{\ell j} = 0\end{math} for all~\begin{math}\ell \neq i\end{math};
		\item	if~\begin{math}i < 0 \end{math}, then~\begin{math}a_{\ell j} = 0\end{math},~\begin{math} 0 \leq \ell \leq k-i \end{math}; and
		\item if~\begin{math}i >  0 \end{math}, then~\begin{math}a_{-\ell j} = 0\end{math},~\begin{math}0 \leq \ell \leq k-i \end{math}.
	\end{itemize}
\end{obs}

We now construct an oriented graph, $\mathcal{G}_{k,  t}$, which is 
a universal target for the family of $k$-dipath $t$-colourable oriented graphs of directed girth at least $k+1$. 
The vertex set~\begin{math}V(\mathcal{G}_{k,  t}) = \mathcal{A}_{k, t}\end{math}.
We describe the edge set informally at first, and then formally.
Suppose the matrices~\begin{math}A = A_{x, c}\end{math} and~\begin{math}B = B_{y, c^\prime}\end{math} are vertices of $\mathcal{G}_{k,t}$.
Let~\begin{math}c(x) = m_1\end{math} and~\begin{math}c^\prime(y) = m_2\end{math}.
Then~\begin{math}AB \in E(G)\end{math} if
\begin{enumerate}
\item $A$ encodes the fact that $x$ has an out-neighbour of colour $m_2$;
\item $B$ encodes the fact that $y$ has an in-neighbour of colour $m_1$;
\item if $A$ encodes the fact that a vertex of colour $m_3$ is joined to $x$ by a directed path of length~\begin{math}i < k\end{math},
then $B$ must encode the fact that a vertex of colour $m_3$ is joined to $y$ by a directed path of length~\begin{math}i+1\end{math};
\item if $B$ encodes the fact that $y$ is joined to a vertex of colour $m_3$ by a directed path of length~\begin{math}i < k\end{math},
then $A$ must encode the fact that $x$ is joined to a vertex of colour $m_3$ by a directed path of length~\begin{math}i + 1\end{math}.
\end{enumerate}

We now formally define the edge set of $\mathcal{G}_{k,  t}$.
Let~\begin{math}A = [a_{ij}]\end{math} and~\begin{math}B = [b_{ij}]\end{math} be vertices of $\mathcal{G}_{k, t}$ (\textit{i.e.}, matrices in $\mathcal{A}_{k,  t}$).
Then~\begin{math}AB \in E(\mathcal{G}_{k,  t})\end{math} if, whenever~\begin{math}a_{0 m_1} = b_{0 m_2} = 1\end{math}, the following conditions all hold:

	\begin{enumerate}
		\item $a_{1 m_2} = 1$;  
		\item $b_{-1 m_1} = 1$; 
		\item  if $0 < i < k-1$ and $a_{-i m_3} = 1$, then  $b_{-(i-1) m_3} = 1$; and  
		\item if $0 < i < k-1$ and $b_{i m_3} = 1$, then  $b_{(i+1) m_3} = 1$. 

	\end{enumerate}

We now establish some properties of $\mathcal{G}_{k,  t}$.

\begin{lem} \label{kdipath:lem2}
	The digraph $\mathcal{G}_{k,  t}$ has directed girth at least $k+1$.  In particular, it is an oriented graph.
\end{lem}

\begin{proof}
	Let~\begin{math}A_1,A_2,\dots,A_\ell, A_1\end{math} be a directed cycle in $\mathcal{G}_{k,  t}$.  Suppose~\begin{math}\ell \leq k\end{math}, and $A_1$ has a $1$ in entry~\begin{math}(0,m_1)\end{math}. This implies  $A_2$ has a $1$ in entry~\begin{math}(-1, m_1)\end{math} and a $1$ in entry~\begin{math}((\ell-1), m_1)\end{math},  contrary to Observation \ref{kdipath:colstruc}.
\end{proof}

\begin{lem} \label{kdipath:lem1}
	For integers $t\geq 2$ and $k \geq 2$ the $k$-dipath chromatic number of $\mathcal{G}_{k, t}$ is at most $t$.
	\label{AtMostT}
\end{lem}		

\begin{proof}
Consider the colouring, $c$, given by $c(A) = m_1$, 
where $m_1$ the is unique column of the matrix of $A$ for which the entry $(0,m_1)$ of $A$ is a $1$. 
We claim that $c$ is a $k$-dipath colouring of $\mathcal{G}_{k,t}$. 
Let~\begin{math} A_1A_2,\dots,A_\ell\end{math} be a directed path of length~\begin{math}\ell \leq k\end{math} in $\mathcal{G}_{k,  t}$. 
If there exists a pair of indices $i$ and $j$ such that~\begin{math}1 \leq i < j \leq k \end{math} and~\begin{math}c(A_i) =c(A_j)\end{math}, 
then  $A_{i+1}$ has a $1$ in entry~\begin{math}(-1, c(A_i))\end{math} and a $1$ in entry~\begin{math}((j-(i+1)), c(A_i))\end{math}, contrary to Observation \ref{kdipath:colstruc}.
\end{proof}

\begin{thm}
	Let~\begin{math}t\geq 2\end{math} and~\begin{math}k\geq 2\end{math} be integers. If $G$ is an oriented graph with directed girth at least $k+1$, then~\begin{math} \chi_{k\text{-dip}}(G) \leq t \end{math} if and only if~\begin{math}G \rightarrow \mathcal{G}_{k,  t}\end{math}.
\label{HomModel}
\end{thm}

\begin{proof}
	Let $G$ be an oriented graph with directed girth at least $k+1$. If~\begin{math}G \rightarrow \mathcal{G}_{k,  t}\end{math}, then by Lemmas \ref{kdipath:lem2} and \ref{kdipath:lem1}, and by Theorem \ref{kdipath:homgirth},~\begin{math} \chi_{k\text{-dip}}(G) \leq t \end{math}. 
	
Suppose ~\begin{math} \chi_{k\text{-dip}}(G) \leq t \end{math}. 
Let $c$ be a $k$-dipath colouring of $G$ using $t$ colours. 
Consider the mapping~\begin{math}\phi: V(G) \rightarrow V(\mathcal{G}_{k,  t})\end{math}, 
where for all~\begin{math}v \in V(G)\end{math},~\begin{math}\phi(v)=A_v\end{math}, the $k$-dipath colouring matrix of $v$ with respect to $c$.  
Let $uv$ be an arc of $G$. 
We claim  $A_uA_v$ is an arc of $\mathcal{G}_{k,  t}$. 
Suppose  $A_u$ has a $1$ in entry~\begin{math}(0,m_1)\end{math}, and $A_v$ has a $1$ in entry~\begin{math}(0,m_2)\end{math} (\textit{i.e.},~\begin{math}c(u) = m_1\end{math} and~\begin{math}c(v) = m_2\end{math}). 
To show that  $A_u$ and $A_v$ are adjacent in $\mathcal{G}_{k,  t}$ we must check that the four conditions 
in the definition are satisfied:
\begin{enumerate}
	\item Since~\begin{math}c(v) = m_2\end{math}, $A_u$ has a $1$ in entry~\begin{math}(1, m_2)\end{math}.
	\item Since~\begin{math}c(u) = m_1\end{math}, $A_v$ has a $1$ in entry~\begin{math}(-1, m_1)\end{math}.
	\item Suppose there exists $i>0$ such that $A_u$ has a $1$ in entry~\begin{math}(-i, m_3)\end{math}. 
	Since $c$ is a $k$-dipath colouring the entries of $A_u$ in column $m_3$ and rows~\begin{math} 0, 1, \ldots, k-i\end{math} are all equal to 0.
	\item Suppose there exists $i>0$ such that $A_v$ has a $1$ in entry $(i, m_3)$. 
	Since $c$ is a $k$-dipath colouring, the entries of $A_v$ in column $m_3$ and rows~\begin{math}-(k-i), -(k-i+1), \ldots, -1, 0 \end{math} are all equal to zero.
\end{enumerate}
This proves the claim.  Therefore~\begin{math}\phi:G \rightarrow \mathcal{G}_{k,  t} \end{math} is a homomorphism.
\end{proof}

\begin{cor}
For  integers $t\geq 2$ and $k\geq 2$,~\begin{math}\chi_{k\text{-dip}}(\mathcal{G}_{k, t}) = t \end{math}.
\end{cor}

\begin{proof}
We show ~\begin{math}\chi_{k\text{-dip}}(\mathcal{G}_{k, t}) \geq t \end{math}.  The result then follows from Lemma \ref{AtMostT}.
Consider the transitive tournament  $T_t$.  
Assigning each vertex a different colour is a $k$-dipath $t$-colouring of $T_t$.
Hence, by Theorem \ref{HomModel}, there is a homomorphism~\begin{math}T_t \to \mathcal{G}_{k,  t}\end{math}.

Since $T_t$ does not have a $k$-dipath $m$-colouring for~\begin{math}m < t \end{math}, by Theorem \ref{HomModel}, there is
no homomorphism~\begin{math}T_t \to \mathcal{G}_{k, m}\end{math}.  
Since a composition of homomorphisms is a homomorphism, it follows that 
\begin{math}\mathcal{G}_{k, m} \not\to \mathcal{G}_{k, t}$ when $m < t\end{math}.
Therefore,~\begin{math}\chi_{k\text{-dip}}(\mathcal{G}_{k, t}) > t-1\end{math}.  
The result now follows.
\end{proof}

Our final result of this section shows that the homomorphism model captures three facts.
First, if~\begin{math}t < k \end{math}, then no oriented graph with directed girth at least $k+1$ and a $k$-dipath $t$-colouring 
can have a directed path of length greater than $t$.  Any such digraph is $t$-dipath $t$-colourable (as the proof shows).
Second, if~\begin{math}t \geq k \end{math}, then any $k$-dipath $t$-colouring of an oriented graph is also a $k$-dipath $(t+1)$-colouring, 
and there exist oriented graphs with $k$-dipath chromatic number $t+1$.
Third, every $(k+1)$-dipath $t$-colouring of an oriented graph is a $k$-dipath $t$-colouring.

\begin{thm}
Let ~\begin{math}k \geq 2\end{math} and~\begin{math}t \geq 2\end{math} be integers.
Then
\begin{enumerate}
\item if $t \leq k$, then~\begin{math}\mathcal{G}_{k, t} \to \mathcal{G}_{t, t}\end{math};
\item~\begin{math}\mathcal{G}_{k, t} \to \mathcal{G}_{k, (t+1)}\end{math} and~\begin{math} \mathcal{G}_{k, (t+1)} \not\to \mathcal{G}_{k, t} \end{math};
\item if $t > k$, then~\begin{math}\mathcal{G}_{(k+1), t} \to \mathcal{G}_{k, t}\end{math} and~\begin{math}\mathcal{G}_{k, t} \not\to \mathcal{G}_{(k+1), t} \end{math}.
\end{enumerate}\label{TargetHoms}
\end{thm}

\begin{proof}
We first prove (1).
Suppose~\begin{math}t \leq k\end{math}.  
Then no oriented graph with directed girth at least $k+1$ and a $k$-dipath $t$-colouring has a 
directed path of length greater than $t$.  
By Theorem \ref{kdipath:acyclicmsw} there is a homomorphism of $G$ to the transitive tournament $T_t$.
In particular,~\begin{math}\mathcal{G}_{k, t} \to T_t \end{math}.
But, by Theorem \ref{HomModel},~\begin{math}T_t \to \mathcal{G}_{t, t}\end{math}.
Therefore,~\begin{math}\mathcal{G}_{k, t} \to \mathcal{G}_{t, t}\end{math}.

We now prove (2).
It is clear that the subgraph of $\mathcal{G}_{k, (t+1)}$ induced by the vertices (colouring matrices) in which 
every entry in column $t+1$ is zero (\textit{i.e.}, $k$-dipath $(t+1)$-colourings in which colour $t+1$ is never used) is
isomorphic to $\mathcal{G}_{k, t}$.  
Thus, $\mathcal{G}_{k, t} \to \mathcal{G}_{k, (t+1)}$.
On the other hand, the transitive tournament on $t+1$ vertices has a homomorphism to 
$\mathcal{G}_{k, (t+1)}$ but not to $\mathcal{G}_{k, t}$.
Therefore, ~\begin{math} \mathcal{G}_{k, (t+1)} \not\to \mathcal{G}_{k, t} \end{math}.

Finally, we prove (3).
Suppose $t > k$.  
Since a $(k+1)$-dipath $t$-colouring is a $k$-dipath $t$-colouring, it follows from Theorem \ref{HomModel} that
\begin{math}\mathcal{G}_{(k+1), t} \to \mathcal{G}_{k, t}\end{math}.
To see the second statement, note that a directed cycle of length $k+1$ has a homomorphism to $\mathcal{G}_{k, t}$ but,
by Lemma \ref{kdipath:lem2}, no homomorphism to $\mathcal{G}_{(k+1), t}$.
Therefore,~\begin{math}\mathcal{G}_{k, t} \not\to \mathcal{G}_{(k+1), t}\end{math}.
\end{proof}

Directed graphs, $G$ and $H$ are called \emph{homomorphically equivalent} if there are
homomorphisms $G \to H$ and $H \to G$.  
If $G$ and $H$ are homomorphically equivalent, then a directed graph admits a 
homomorphism to $G$ if and only if it admits a homomorphism to $H$.
A directed graph is a \emph{core} if it is not homomorphically equivalent to any proper subgraphs.
Every directed graph $G$ has a minimal subgraph $H$ which is a core, and to which $G$ is homomorphically equivalent;  $H$ is unique up to isomorphism, and is called \emph{the core of $G$} (see \cite{F82} and \cite{W82}).  
We note that the core, $H$, of $G$ is an induced subgraph of $G$ because, by minimality,
any homomorphism~\begin{math}G \to H\end{math} must map $H$ onto itself.

\begin{cor}
Let $t$ and $k$ be positive integers. If $t \leq k$, then the core of $\mathcal{G}_{k, t}$ is $T_t$.
\label{Core}
\end{cor}
\begin{proof}
Note that $T_t$ is a core. The existence of the required homomorphism is noted in the proof of statement (1) in Theorem \ref{TargetHoms}.
\end{proof}

\section{Complexity of $k$-dipath Colourings} \label{kdipath:comsec}

In this section we consider the following decision problem.

\noindent \underline{$k$-DIPATH $t$-COLOURING}\\
Instance: an oriented graph, $G$.\\
Question: does $G$ have a $k$-dipath $t$-colouring?

The dichotomy theorem stated below covers the cases where $k = 2$.
We shall find a similar result for all remaining cases.

\begin{thm}  [\cite{MAYO12, KMY09}]
	Let $t \geq 1$ be a fixed integer. If~\begin{math} t \leq 2 \end{math}, then $2$-DIPATH $t$-COLOURING is Polynomial. If~\begin{math} t \geq 3 \end{math}, then $2$-DIPATH $t$-COLOURING is NP-complete.
\label{2DPDichotomy}
\end{thm}

Given a simple graph $G$, we construct an oriented graph, $H_{k,t}$ (\begin{math}t > k \geq 3\end{math}), such that~\begin{math}\chi_{k\text{-dip}}(H_{k,t}) \leq t\end{math} if and only if~\begin{math}\chi(G) \leq t \end{math}. Let $G$ be a simple graph, and let $\tilde{G}$ be an arbitrary acyclic orientation of $G$. 
Corresponding to each vertex~\begin{math}v \in V(\tilde{G})\end{math} the oriented graph $H_{k,t}$ contains:
\begin{enumerate}
\item vertices $v_\text{in}$ and $v_\text{out}$;
\item a transitive tournament on~\begin{math}t-k+1\end{math} vertices with source vertex $s_v$ and sink vertex $t_v$:
\item vertices~\begin{math}v^\prime_1, v^\prime_2, \ldots, v^\prime_{k-2}\end{math};
\item  the directed path~\begin{math}t_v,v^\prime_1, v^\prime_2, \ldots, v^\prime_{k-2},v_\text{in}\end{math}, 
which has length $k-1$; and
\item an arc $v_\text{out}s_v$.
\end{enumerate}
For each arc~\begin{math}uv \in E(\tilde{G})\end{math}, the oriented graph $H_{k,t}$ is augmented by adding 
the arc~\begin{math}u_\text{out}v_\text{in}\end{math}.

This completes the construction of $H_{k,t}$.  It can clearly be carried out in polynomial time.
We note that since $\tilde{G}$ has no directed cycles, by construction the same is true of $H_{k,t}$.
Hence $H_{k,t}$ (formally) has directed girth at least $k+1$.  
We also note that each vertex $v_\text{out}$ has in-degree 0 and and each vertex 
$v_\text{in}$ has out-degree $0$.

\begin{obs}
	\begin{math}\chi_{k\text{-dip}}(H_{k,t}) \geq t\end{math}.
\end{obs}
\begin{proof}
For any vertex~\begin{math}v \in V(G)\end{math}, observe that the subgraph of $H_{k,t}$ induced by the~\begin{math}t-k+1\end{math} vertices of the transitive tournament corresponding to  $v$, together with the vertices~\begin{math}v^\prime_1, v^\prime_2, \ldots, v^\prime_{k-2},v_\text{in}\end{math}
is a $k$-dipath clique on $t$ vertices.
\end{proof}

\begin{obs} \label{kdipath:obslift2}
Let~\begin{math}t > k \geq 3\end{math}. If $H_{k,t}$ has a $k$-dipath $t$-colouring, then for every $v \in V(G)$ and every $k$-dipath $t$-colouring, $c$, of $H_{k,t}$, ~\begin{math}c(v_\text{out}) =  c(v_\text{in})\end{math}.
\end{obs}
\begin{proof}
For any $v \in V(G)$, observe that the subgraph of $H_{k,t}$ induced by the vertex $v_\text{out}$, the~\begin{math}t-k+1\end{math} vertices of the transitive tournament corresponding to $v$ and the vertices~\begin{math}v^\prime_1, v^\prime_2, \ldots, v^\prime_{k-2}\end{math}  is a $k$-dipath clique on $t$ vertices.
Since the subgraph of $H_{k,t}$ induced by the $t-k+1$ vertices of the transitive tournament corresponding to  $v$, together with the vertices~\begin{math}v^\prime_1, v^\prime_2, \ldots, v^\prime_{k-2}\end{math} and $v_\text{in}$ is also $k$-dipath clique on $t$ vertices, and $c$ is a $k$-dipath $t$-colouring, it follows that~\begin{math}c(v_\text{out}) =  c(v_\text{in})\end{math}.
\end{proof}

\begin{lem} \label{kdipath:nplift2}
If $G$ is a simple graph and $H_{k,t}$ is constructed from $G$ as above, then, for all~\begin{math}
t > k \geq 3
\end{math}, the graph $G$ is $t$-colourable if and only if $H_{k,t}$ has a $k$-dipath $t$-colouring. 
\end{lem}

\begin{proof}
Suppose  $H_{k,t}$ has a $k$-dipath $t$-colouring, $c$.
Let~\begin{math}
\phi: V(G) \rightarrow \{1,2,3, \dots, t\}
\end{math} be defined by~\begin{math}
\phi(v) = c(v_\text{in})
\end{math}.
We claim that $\phi$ is a proper colouring of the graph $G$.
Suppose~\begin{math}
ab \in E(\tilde{G})
\end{math}.
Then, ~\begin{math}
a_\text{out}b_\text{in} \in E(H_{k,t})
\end{math}, and so  
~\begin{math}
 c(a_\text{out}) \neq c(b_\text{in}).
 \end{math}
By Observation~\ref{kdipath:obslift2},~\begin{math}
c(b_\text{out}) = c(b_\text{in}).
\end{math}
Therefore,~\begin{math}
c(a_\text{in}) \neq c(b_\text{in})
\end{math}, and~\begin{math}\phi(a) \neq \phi(b)\end{math}.
This proves the claim.  Hence $G$ is $t$-colourable.

Suppose now that $G$ has a $t$-colouring, $\phi$.	
We construct a $k$-dipath $t$-colouring, $c$, of $H_{k,t}$.
For each vertex~\begin{math}
v \in V(G)
\end{math}, set  ~\begin{math}
c(v_\text{out}) = c(v_\text{in}) = \phi(v)
\end{math}.
We claim that $c$ can be completed to a $k$-dipath $t$-colouring of $H_{k,t}$.

Since, for every~\begin{math}v \in V(G)\end{math}, if~\begin{math}
\phi(v_\text{out})= i
\end{math},  then vertices of the transitive tournament corresponding to $v$  together with the vertices~\begin{math}
v^\prime_1, v^\prime_2, \ldots, v^\prime_{k-2}
\end{math}  can be assigned colours from the set~\begin{math}\{1,2,3 \dots, t\} \setminus \{i\}\end{math} so that the resulting colouring has the property that no two vertices at weak distance at most $k$ are assigned the same colour.  

This proves the claim.
\end{proof}

\begin{thm}
	Let $t$ and~\begin{math}
	k \geq 3
	\end{math} be fixed positive integers. 
	When restricted to instances of directed girth at least~\begin{math}
	k+1
	\end{math}, the decision problem
	\emph{$k$-DIPATH $t$-COLOURING} is NP-complete if~\begin{math}
	t > k
	\end{math}, and Polynomial if ~\begin{math}
	t \leq k
	\end{math}.
	\label{Dichotomy1}
\end{thm}

\begin{proof}
The problem is clearly in NP.  
If~\begin{math}
t > k
\end{math} then NP-completeness follows from Lemma \ref{kdipath:nplift2}.

Suppose~\begin{math}
t \leq k
\end{math}.  
By Corollary \ref{Core}, an oriented graph $G$ with directed girth at least $k+1$ has a $k$-dipath $t$-colouring if
and only if it admits a homomorphism to the transitive tournament $T_t$.
Homomorphism to $T_t$ can be checked in polynomial time as shown by \cite{BHG88}.
The result now follows. 
\end{proof}	

We now show that if the girth restriction is removed, then the dichotomy changes.

\begin{thm}
Let $k$ and $t$ be positive integers.
If~\begin{math} t \leq 2 \end{math}, then \emph{$k$-DIPATH $t$-COLOURING} is Polynomial.
If~\begin{math} t \geq 3 \end{math}, then \emph{$k$-DIPATH $t$-COLOURING} is NP-complete.
\end{thm}
\begin{proof}
Suppose first that~\begin{math} t \leq 2 \end{math}.
If~\begin{math}
k = 1
\end{math}, then the condition that vertices joined by a directed path of length at most $k$ must be assigned different colours 
is the same as the condition that adjacent vertices must be  assigned different colours.
Hence,  a digraph $G$ has a $k$-dipath $t$-colouring if and only if the underlying graph of $G$ has a $t$-colouring.
The latter problem is Polynomial.
If $k=2$,  $k$-dipath $t$-colouring is Polynomial by Theorem \ref{2DPDichotomy}.
For~\begin{math}
k \geq 3
\end{math}, if $G$ has a directed path of length at least 3 or a directed 3-cycle (which is easy to check),
then, since~\begin{math}
t \leq 2
\end{math}, it has no $k$-dipath $t$-colouring.  
Otherwise, ($G$ has no directed path of length greater than 2), 
a $k$-dipath $t$-colouring is a $2$-dipath $t$-colouring, which is Polynomial.

Now suppose~\begin{math}
t \geq 3
\end{math}.
If~\begin{math}
k=2
\end{math}, then $k$-dipath $t$-colouring is NP-complete by Theorem \ref{2DPDichotomy}.
Hence, assume~\begin{math} k \geq 3 \end{math}.
If $t > k$, the result follows from Theorem \ref{Dichotomy1}.
Hence we may also assume $t \leq k$.
Similarly as above,  if $t < k$ then 
no oriented graph with a directed path of 
length greater than $t$ (which is easy to check), can have a
a $k$-dipath $t$-colouring. 
Otherwise ($G$ has no directed path of length greater than $t$), 
a $k$-dipath $t$-colouring is a $t$-dipath $t$-colouring.
Thus, we may further assume that~\begin{math}
t = k \geq 3
\end{math}.

We have previously noted that the problem belongs to NP.
The transformation is from $t$-colouring.
Suppose an instance of $t$-colouring, a simple, undirected graph $G$ is given.
We will replace each edge of $G$ by the oriented graph  which we now construct.

Let $C$ be the directed cycle of length $t$ with vertex sequence~\begin{math}
v_0, v_1, \ldots, v_{t-1}, v_0
\end{math}.
Add four new vertices~\begin{math}
x_0, x_1, y_0, y_1
\end{math}, and arcs to make the directed paths
\begin{math}
x_0, x_1, v_1
\end{math} and~\begin{math}
y_0, y_1, v_2
\end{math}.
This completes the construction of $F$.
Observe that $x_0$ is joined to every vertex of $C$ except $v_0$ by a directed path of length at most $t$.
Every vertex of $C$ is assigned a different colour in a $k$-dipath $t$-colouring of $F$.
Hence, the vertices $x_0$ and $v_0$ must be assigned the same colour.
Similarly, the vertices $y_0$ and $v_1$ must be assigned the same colour.
In particular, $x_0$ and $y_0$ must be assigned different colours.
Furthermore, any assignment of two different colours to $x_0$ and $y_0$ can be extended to a 
$k$-dipath $t$-colouring of $F$.

Construct an oriented graph $G^\prime$ from $G$ as follows. 
Replace each edge $xy$ of $G$ with a copy $F_{xy}$ of $F$ by 
identifying $x_0$ with $x$ and $y_0$ with $y$.
The construction can clearly be carried out in polynomial time.
Observe that each vertex in~\begin{math}
V(G) \cap V(G^\prime)
\end{math} has in-degree 0.

We claim that $G$ is $t$-colourable if and only if $G^\prime$ has a $k$-dipath $t$-colouring.
Suppose that a $t$-colouring of $G$ is given.
For each copy of $F$ in $G^\prime$, this assignment gives different colours to $x_0$ and $y_0$.
By our earlier observation, this assignment can be  extended to a $k$-dipath $t$-colouring of $F$.
Since each vertex in~\begin{math}
V(G) \cap V(G^\prime)
\end{math} has in-degree 0, there is no directed path of positive length 
joining vertices in different copies of $F$, this results in a  $k$-dipath $t$-colouring of $G^\prime$.
Now suppose a $k$-dipath $t$-colouring of $G^\prime$ is given.
By our observation on colourings of $F$, any two vertices which are adjacent in $G$ are assigned 
different colours.  Hence restricting this colouring to $V(G)$ gives a $t$-colouring of $G$.
This completes the proof.
\end{proof}

The problem of deciding whether the $k$-th power of a graph $G$ is $t$-colourable is known to be
NP-complete (see \cite{L95} and \cite{M83}).
The results above imply that, if~\begin{math}
t > k
\end{math}, the problem of deciding whether the underlying graph of the
$k$-th power 
of an oriented graph $G$ (\textit{i.e.}, two vertices are adjacent if and only if they are at weak distance at most $k$) 
is $t$-colourable is NP-complete, even when restricted to powers of oriented graphs with directed girth at least $k+1$,
and also that, if~\begin{math}
t = k \geq 3
\end{math}, 
the problem of deciding whether the underlying graph of the $k$-th power 
of an oriented graph $G$ is $t$-colourable is NP-complete.

\bibliographystyle{abbrvnat}
\bibliography{references}

\end{document}